\documentclass[conference]{IEEEtran}
\IEEEoverridecommandlockouts
\usepackage{cite}
\usepackage{amsmath,amssymb,amsfonts}
\usepackage{algorithmic}
\usepackage{graphicx}
\usepackage{textcomp}
\usepackage{xcolor,balance}
\usepackage[ruled,vlined]{algorithm2e}
\def\BibTeX{{\rm B\kern-.05em{\sc i\kern-.025em b}\kern-.08em
  T\kern-.1667em\lower.7ex\hbox{E}\kern-.125emX}}

\newcommand{\V}[2][]{\mathbb{V}\mathrm{ar}_{#1}\!\left[{#2}\right]}
\newcommand{\E}[2][]{\mathbb{E}_{#1}\!\left[{#2}\right]}
\newcommand{\tr}[1]{\mathrm{tr}\!\left[{#1}\right]}

\begin{document}

\title{Dual-Function Radar-Communication 
	System Aided by Intelligent Reflecting Surfaces\\
}

\author{\IEEEauthorblockN{Yikai Li}
\IEEEauthorblockA{\textit{School of Computing} \\
\textit{Southern Illinois University}\\
Carbondale, IL, USA \\
yikai.li@siu.edu}
\and
\IEEEauthorblockN{Athina Petropulu}
\IEEEauthorblockA{\textit{Dept. of Electrical and Computer Engineering} \\
\textit{Rutgers University}\\
Piscataway, NJ, USA \\
athinap@rutgers.edu}
}

\maketitle
\thispagestyle{plain} 
\pagestyle{plain}    

\begin{abstract}
We propose a novel design of a dual- function radar- communication (DFRC) system aided by an Intelligent Reflecting Surface (IRS). We consider a scenario with one target and multiple communication receivers,  where there is no line-of-sight between the radar and the target.
 The radar precoding  matrix and the IRS weights are optimally designed to maximize the weighted sum of the  signal-to-noise ratio (SNR)  at the radar receiver and the SNR at the  communication receivers subject to power constraints and   constant modulus constraints on the IRS weights. 
The problem is decoupled into two sub-problems, namely, waveform design and IRS weight design, and is solved via alternating optimization. The former subproblem is solved via  linear programming, and the latter via manifold optimization with a quartic polynomial objective. 
The
key contribution of this paper lies in  solving the IRS weight design sub-problem that is based on the optimization of a quartic objective function  in the IRS weights, and is subject to  unit modulus-constraint on the IRS weights. Simulation results are provided to show the  
convergence behavior of the proposed algorithm under different system configurations, and the effectiveness of using IRS to  improve radar and communication performance.
\end{abstract}


\vspace{-2mm}
\section{Introduction}
\vspace{-1mm}

Dual-function radar-communication (DFRC) systems
offer sensing and communication functionality out of a single platform and using the same waveform. As such, they efficiently use the spectrum and also offer smaller device size and lower hardware cost as compared to systems that perform sensing and communication out of separate platforms \cite{Liu2020}.
Several works have focused on the design of DFRC systems that are either 
radar-centric, or communication-centric, and have considered joint designs \cite{Zhang2020d}, \cite{Xu2021} and \cite{Liu2018}.

Intelligent reflecting surface (IRS) technology has been gaining a lot of ground for the design of next generation communication systems \cite{Wu2020}.
Low-cost IRS platforms can be deployed in the path of the signal to change the signal's propagation characteristics. By  controlling the amplitudes and/or phases of the responses of the individual IRS elements, the IRS can  implement fine-grained passive beamforming for directional signal nulling or enhancement  \cite{Wu2019}.

While there is significant body of literature on reaping the benefits of IRS for  wireless communication systems  \cite{Wu2020,Wu2019}, the use of IRS in  radar is only now starting to receive attention \cite{Buzzi2021,Lu2021,Lu2021a,Wang2021a,Aubry2021}. 
A smart IRS can enhance the power of the reflected target signal, or the signal-to-noise ratio (SNR) at the receive array, and thus  improve the  target detection performance \cite{Buzzi2021,Lu2021,Lu2021a}.
In \cite{Buzzi2021}, the radar  transmits/receives through two separate beams,  one pointing towards the search
direction and the other towards  the IRS. The smart IRS can be controlled to focus the impinging wavefront towards the radar during the reception stage, or  towards the target during the transmission stage. The IRS can also be placed near the radar to fill the area covered by the 
radar beam, which essentially plays the role of a feed antenna \cite{Buzzi2021}. A radar with a far-away placed IRS can form a low-cost bistatic radar system, offering additional  diversity gain  \cite{Buzzi2021}. Using multiple distributed IRS platforms can lead to 
sharper main lobe and more accurate angle estimation, better angle resolution, and higher detection probability \cite{Lu2021,Lu2021a}.  The IRS can also help mitigate the interference between spectrum sharing systems \cite{Wang2021a},
or used for non-line of sight (N-LOS) radar surveillance \cite{Aubry2021,Wang2021b}. 
{Each IRS element induces a  phase shift $\phi$ on the incident signal, thus the responses of the elements, or  IRS weights,  have unit magnitude.}

Although the effectiveness of IRS on independent  radar or communication systems has been studied,  research on the effectiveness of IRS in integrated DFRC systems is still in  early stages \cite{Wang2021,Jiang2021}. In \cite{Wang2021}, the design of an IRS-aided DFRC is based on jointly minimizing the  radar beampattern error and the communication multi-user interference
{by alternatively optimizing the   radar precoding matrix and IRS weights.}
In that case,  the objective function is a quadratic polynomial {in the IRS weight matrix}. 
{
In \cite{Jiang2021}, the design is based on  radar SNR maximization subject to a communication receiver SNR constraint. The objective function of \cite{Jiang2021} is quartic in the IRS weight matrix.
The IRS  design problem is relaxed 
into a semi-definite programming (SDP) problem {by dropping the unit modulus IRS weight  constraint}, and is then  solved using semi-definite relaxation (SDR) techniques.} 
However,  a rank-one solution is hard to obtain and the algorithm convergence cannot be guaranteed with the Gaussian randomization approximation method \cite{Jiang2021}.

In this paper we consider an IRS aided DFRC system, for a scenario in which 
there is one target and multiple communication receivers. There is no line-of-sight between the radar and the target, and target detection is carried out based on signals reflected by the IRS.
We  propose a novel co-design of  IRS weights and  radar waveform  for  maximizing the weighted sum of the received SNR at the radar  and communication receivers based on constant modulus constraints on the IRS weights. 
The  optimization problem is formulated as  alternating optimization of  two sub-problems, namely, radar waveform design and IRS weight design. The first problem is solved via linear programming, and the second  via manifold optimization with a quartic polynomial objective function {with respect to the IRS weights}. 
Derivation of the gradient of the quartic objective function  is more challenging than that of {a typical} quadratic objective function. For this we borrow  ideas from the work in \cite{Alhujaili2020}, which focuses on a different scenario, namely,  a  disturbance power minimization problem at
	the output of the matched filter in a single antenna cognitive radar.
		Our proposed method allows us to deal with both the unit modulus IRS weight constraint and a quartic objective function, and this is something that has not been explored in 
		\cite{Jiang2021,Wang2021}.

	

{The closest  work on DFRC to this paper is \cite{Jiang2021}, which also considers improving the SNR for the radar and communication receivers under a radar transmit power budget. }
{
However, in \cite{Jiang2021}, the highly non-convex constant-modulus constraint for the IRS weights is relaxed during the optimization process,   leading to less accurate solution and slower algorithm convergence. 
}  
{Due to the constant-modulus constraint, the solution is constrained onto a complex circle manifold, and thus solutions obtained in Euclidean space will  deviate from that manifold. The work of \cite{Jiang2021} uses Gaussian randomization, i.e., it  relaxes the constraint and then generates randomized solutions and picks the one which maximizes the objective function. However, this approach  is not efficient since it requires a large number of realizations to get an accurate enough solution.}
The work of \cite{Wang2021} is also related to our work, except that  
 {the design criterion is different, leading to a different optimization problem.}

\textit{\textbf{Notation:}}      $\mathbf M^{H}$, $\mathbf M^{\ast}$,  and
$\mathbf M_{k,l}$ are  the conjugate-transpose, conjugate,  and the  $(k,l)$-th  element  of   a  matrix $\mathbf M$, respectively.
$\E[]{X}$ and $\V{X}$ are the  mean and variance   of    a random variable $X$,  respectively. $\tr{\mathbf M}$ denotes the trace of a square matrix $\mathbf M$. In addition, $\mathbf 0_{m\times n}$ and $\mathbf I_{m}$ respectively denote an $m \times n$ matrix with all zero elements and an $m \times m$ identity matrix. Moreover, $\otimes$ and $\circ$ are respectively the Kronecker product operator and  Hadamard product operator. {$\mathbf v \sim \mathcal{CN}(\mathbf 0_{m \times 1}, \sigma^2 \mathbf I_m)$ represents that $\mathbf v$ is an $m \times 1$ circularly symmetric complex Gaussian vector with zero mean and covariance matrix $\sigma^2 \mathbf I_m$.}

\vspace{-2mm}
\section{System model}
\vspace{-1mm}
\label{sec:models}

\begin{figure}[!t]\vspace{0mm} 
	\hspace{3mm} 
	\def\svgwidth{220pt} 
	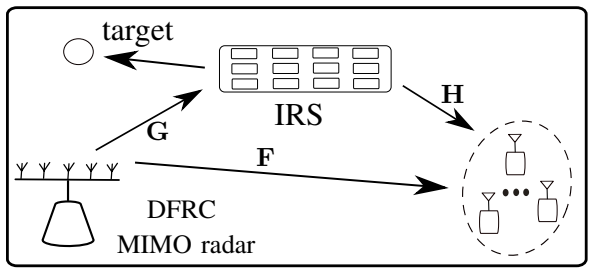 \vspace{-3mm}
	\caption{Intelligent 
		reflecting surface assisted DFRC system.}\vspace{-8mm} \label{fig:system_model}
\end{figure}

Let us consider an IRS-aided DFRC system (Fig. \ref{fig:system_model}), consisting of a MIMO radar with an $M$-antenna uniform linear array (ULA) as transmitter,  and an $M$-antenna ULA receive array, both with antenna spacing $d$.  
There are  $K$ single-antenna communication receivers and one point target.
The DFRC is aided by  an IRS platform,   modeled as an $N$-element uniform planar array (UPA), 
with $N_x$ elements per row and  $N_y$ per column.
{We assume that there exists no line of sight (LOS) between the radar and target.} All channels are assumed to be flat fading and  perfectly known.

The signal transmitted by the radar can be written as

\vspace{-7mm}
\begin{eqnarray}
	\mathbf X = \mathbf W \mathbf s,
\end{eqnarray}
\vspace{-7mm}

\noindent where $\mathbf W \in \mathbb C^{M \times M}$ and $\mathbf s \in \mathbb C^{M \times 1}$ respectively represent the radar precoding matrix and signal vector intended for the  communication users. 
{Radar transmit signal which is reflected by the IRS, then by the target, and the  target echo arrives at the radar receiver  via the IRS (since there is no direct path between the radar and the target).}
The received signal equals

\vspace{-6mm}
\begin{eqnarray}
	\!\!\!\!\!\!\!\!\!\!\!\!&&\mathbf r_r = \eta \mathbf G^T\mathbf \Theta \mathbf a_R(\phi_h,\phi_v) \mathbf a_R^T(\phi_h,\phi_v) \mathbf \Theta \mathbf G  \mathbf W \mathbf s + \mathbf n_r \nonumber \\&&=\mathbf F_r \mathbf W \mathbf s + \mathbf n_r, \label{(2)}
\end{eqnarray}
\vspace{-6mm}

\noindent {where $\eta$ is the complex  channel coefficient corresponding to  the radar-IRS-target-IRS-radar path; } $\mathbf a_R(\phi_h,\phi_v) = \mathbf a_{R_y}(\phi_h,\phi_v) \otimes \mathbf a_{R_x}(\phi_h,\phi_v) $ is the steering vector of the IRS where
$
\mathbf a_{R_y}(\phi_h,\phi_v) = [1,\; e^{j 2 \pi d \cos{\phi_h}\sin{\phi_v}/\lambda},\;\cdots,\;\\e^{j 2 \pi (N_y-1) d \cos{\phi_h}\sin{\phi_v}/\lambda}]^T
$ and $\mathbf a_{R_x}(\phi_h,\phi_v)$ similarly defined based on $N_x$;
$\phi_h$ and $\phi_v$ are respectively the angles of the target relative the IRS in the azimuth and elevation planes; $\mathbf \Theta = \text{diag}([e^{j \theta_1},\cdots,e^{j \theta_N}])$ is the phase shift matrix of the IRS, $\theta_n$ is the delivered phase shift of the $n$-th IRS element for $n \in \{1, \cdots, N\}$ and $N = N_x N_y$ is the total number of IRS elements; 
$\mathbf G$ is the  channel between the radar and the IRS elements, {and $\mathbf n_r \sim \mathcal{CN}(\mathbf 0_{M \times 1}, \sigma_r^2 \mathbf I_M)$ models the additive white Gaussian noise (AWGN) at the radar receiver where $\sigma_r^2$ is the average noise power per radar receiver antenna.}

The signal reaches the communication receivers through a direct path  also through  signal reflected by the IRS. 
Received signal at the communication receivers can be written as

\vspace{-6mm}
\begin{eqnarray}
	\mathbf r_c = \left(\mathbf F+ \mathbf H \mathbf \Theta \mathbf G\right) \mathbf W \mathbf s + \mathbf n_c = \mathbf F_c \mathbf W \mathbf s + \mathbf n_c,
\end{eqnarray}
\vspace{-6mm}

\noindent where $\mathbf F$ is   channel between the radar and the  communication receivers, and  $\mathbf H$  channel  between the IRS and the communication receivers, {and $\mathbf n_c \sim \mathcal{CN}(\mathbf 0_{K \times 1}, \sigma_c^2 \mathbf I_K)$ models the AWGN at the communication receivers where $\sigma_c^2$ is the average noise power per single-antenna communication receiver.}

The output SNRs at the radar and communication receivers can be respectively written as

\vspace{-6mm}
\begin{eqnarray}
	&&\!\!\!\!\!\!\!\!\!\!\!\!\!\!\!\gamma_r = \E{\tr{\mathbf r_r \mathbf r_r^H}} / \sigma_r^2 = \tr{\mathbf F_r \mathbf W \mathbf W^H \mathbf F_r^H} / \sigma_r^2, \\&&\!\!\!\!\!\!\!\!\!\!\!\!\!\!\!\gamma_c = \E{\tr{\mathbf r_c \mathbf r_c^H}}/ \sigma_c^2 = \tr{\mathbf F_c \mathbf W \mathbf W^H \mathbf F_c^H} / \sigma_c^2.
\end{eqnarray}
\vspace{-8mm}

\section{System Design}
\vspace{-1mm}
\label{sec:pagestyle}
We will design  the  precoder matrix $\mathbf W$, and the IRS phase shift matrix $\mathbf \Theta$, so that we optimize  a weighted  combination of the  radar received SNR and that delivered at the communication system, with $\alpha$ being a weight, i.e., 
\begin{subequations} \label{eqn:orig_prob}
	\begin{eqnarray} 
		\max_{\mathbf W, \mathbf \Theta}&&	\;\;\;\;	(1-\alpha)\gamma_r + \alpha \gamma_c\\ \mathrm{s.t.} && \;\;\;\; |\mathbf \Theta_{n,n}| = 1,\;\; \forall n \in \{1,\cdots,N\}\label{eqn:unit_modu}\\&& \;\;\;\; \tr{\mathbf W \mathbf W^H} = P_0 \label{eqn:tot_power}\\&& \;\;\;\; \|\mathbf W \mathbf W^H-\mathbf R_d\| \leq \gamma_{bp} \label{eqn:beam_pattern_cons}
	\end{eqnarray}
\end{subequations}
\vspace{-6mm}

\noindent where (\ref{eqn:unit_modu}) is the unit modulus constraint for the IRS phase shift matrix $\mathbf \Theta$ and  $\mathbf \Theta_{n,n}$ denotes the $n$-th diagonal element of the square matrix $\mathbf \Theta$, (\ref{eqn:tot_power}) represents the total transmit power constraint at the radar,  $P_0$ is the radar power budget, and (\ref{eqn:beam_pattern_cons}) indicates that the beam pattern deviation from a desired one should be within  $\gamma_{bp}$ from a  pre-defined threshold. Here, $\mathbf R_d$ is the covariance matrix of the desired waveform.

The  optimization problem of (\ref{eqn:orig_prob}) is highly non-convex. However, it can be efficiently solved via decomposed into two sub-problems, to be solved in  an alternating way. Specifically, the first sub-problem can be defined as maximizing the objective function by solving for the precoder matrix $\mathbf W$, taking the IRS phase shift matrix $\mathbf \Theta$ as constant. The second problem can be defined as solving for $\mathbf \Theta$ by taking $\mathbf W$ as  constant. These two sub-problems are being solved alternatingly until the objective function converges. \\

\vspace{-4mm}
\noindent \underline{Sub-problem 1.} 
The first sub-problem optimizes the objective function with respect to $\mathbf W$ for  fixed  $\mathbf \Theta$. The objective function can be written as 
$f(\mathbf W) =  (1- \alpha)\tr{ \mathbf W \mathbf W^H \mathbf F_r^H \mathbf F_r}/ \sigma_r^2 + \alpha \tr{ \mathbf W \mathbf W^H \mathbf F_c^H \mathbf F_c}/ \sigma_c^2 =  \tr{\mathbf W \mathbf W^H \mathbf C},$
where $\mathbf C = (1- \alpha) \mathbf F_r^H \mathbf F_r/ \sigma_r^2 + \alpha\mathbf F_c^H \mathbf F_c/ \sigma_c^2$. Thereby, the sub-problem 1 can be written as

\vspace{-6mm}
\begin{small}
\begin{subequations} \label{eqn:sub_prob_1}
	\begin{eqnarray} 
		\max_{\mathbf W}&&	\;\;\;\;	\tr{\mathbf W \mathbf W^H \mathbf C}\\  \mathrm{s.t.}&& \;\;\;\; \tr{\mathbf W \mathbf W^H} = P_0 \label{eqn:}\\&& \;\;\;\; \|\mathbf W \mathbf W^H-\mathbf R_d\| \leq \gamma_{bp} \label{eqn:}
	\end{eqnarray}
\end{subequations}
\end{small}
\vspace{-6mm}

\noindent This is a linear programming problem with variable  $\mathbf W \mathbf W^H$. Then, the precoder matrix $\mathbf W$ can be computed as the square root matrix of $\mathbf W \mathbf W^H$.\\

\vspace{-4mm}
\noindent \underline{Sub-problem 2.}
In the second sub-problem, we use the value of $\mathbf W$ obtained in sub-problem 1, and optimize with respect to $\mathbf \Theta$. The objective function $f = \tr{\mathbf W \mathbf W^H \mathbf C}$  can be re-written as a function of $\mathbf \Theta$ by expressing ${\bf C}$ in terms of  
$\mathbf F_r= \eta \mathbf G^T\mathbf \Theta \mathbf a_R(\phi_h,\phi_v) \mathbf a_R^T(\phi_h,\phi_v) \mathbf \Theta \mathbf G $
and $\mathbf F_c=\mathbf F+ \mathbf H \mathbf \Theta \mathbf G$, i.e., 
\vspace{-6mm}
\begin{eqnarray}
	&&f{(\mathbf \Theta)} = t_4+t_2+t_1+t_0,  
\end{eqnarray}
\vspace{-6mm}

\noindent where $t_4$, $t_2$, $t_1$ and $t_0$ are respectively the quartic, quadratic, linear and constant terms with regard to $\mathbf \Theta$, defined as 

\vspace{-4mm}
\begin{small}
	\begin{subequations} \label{eqn:}
		\begin{eqnarray} 	\!\!\!\!\!\!\!\!\!\!\!\!\!\!\!\!\!\!\!\!\!\!\!\!&&t_4 = (1-\alpha) \eta^2 \tr{\mathbf G^T \mathbf \Theta \mathbf R \mathbf \Theta \mathbf G \mathbf W \mathbf W^{\!H}\! \mathbf G^{\!H}\! \mathbf \Theta^H\! \mathbf R^H\! \mathbf \Theta^H\! \mathbf G^{\ast}}/\sigma_r^2 ,\\\!\!\!\!\!\!\!\!\!\!\!\!\!\!\!\!\!\!\!\!\!\!\!\!&&t_2 =   \alpha  \tr{\mathbf W \mathbf W^H \mathbf G^H \mathbf \Theta^H \mathbf H^H \mathbf H \mathbf \Theta \mathbf G}\!/\sigma_c^2 ,\\\!\!\!\!\!\!\!\!\!\!\!\!\!\!\!\!\!\!\!\!\!\!\!\!&& t_1 = \alpha\! \left( \!\tr{\mathbf W \mathbf W^H \!\mathbf G^H \!\mathbf \Theta^H\! \mathbf H^H\! \mathbf F} \!+\!   \tr{\mathbf W \mathbf W^H \mathbf F^H \mathbf H \mathbf \Theta \mathbf G}\!\right)\!/\sigma_c^2, 	\\\!\!\!\!\!\!\!\!\!\!\!\!\!\!\!\!\!\!\!\!\!\!\!\!&& t_0 = \alpha  \tr{\mathbf W \mathbf W^H \mathbf F^H \mathbf F}/\sigma_c^2,
		\end{eqnarray}
	\end{subequations}
\end{small}
\vspace{-5mm}

\noindent where 
$\mathbf R = \mathbf a_R(\phi_h,\phi_v) \mathbf a_R^T(\phi_h,\phi_v) $.

The term $t_0$ is not a function of $\mathbf \Theta$, thus  can be safely discarded. Thereby, sub-problem 2 can be re-formulated as 
\vspace{-2mm}
\begin{subequations} \label{eqn:sub_prob_2}
	\begin{eqnarray} 
		\!\!\!\!\!\!\!\!\!\max_{\mathbf \Theta}&&	\!\!\!\!\!	t_4+t_2+t_1\\ \!\!\!\!\!\!\!\!\! \mathrm{s.t.}&& \!\!\!\!\! |\mathbf \Theta_{n,n}| = 1,\;\; \forall n \in \{1,\cdots,N\}
	\end{eqnarray}
\end{subequations}
\vspace{-6mm}

To make the objective function more mathematical tractable, the quartic term $t_4$ can be re-written in vector form as

\vspace{-5mm}
\begin{small}
\begin{eqnarray}
	t_4 = (1-\alpha) (\eta^2 / \sigma_r^2) \sum_{i=1}^{M} \sum_{j=1}^{M} h_{ij}(\boldsymbol{\theta}) h_{ij}^{\ast}(\boldsymbol{\theta}),
\end{eqnarray}
\end{small}
\vspace{-3mm}

\noindent where $h_{ij}(\boldsymbol{\theta}) = \boldsymbol{\theta}^T [\mathbf R \circ (\mathbf G \mathbf w_j \mathbf g_i^T)^T] \boldsymbol{\theta}$, $\boldsymbol{\theta}=\mathbf\Theta \mathbf 1_{N \times 1}$, $\mathbf w_j$ is the $j$-th column of $\mathbf W$, and $\mathbf g_i$ is the $i$-th column of $\mathbf G$.

\noindent Meanwhile the quadratic term $t_2$ can be re-arranged as

\vspace{-6mm}
\begin{eqnarray}
	t_2 = \boldsymbol{\theta}^H \mathbf D_1 \boldsymbol{\theta},
\end{eqnarray}
\vspace{-6mm}

\noindent where  $\mathbf D_1= (\alpha / \sigma_c^2)(\mathbf H^H \mathbf H) \circ (\mathbf G \mathbf W \mathbf W^H \mathbf G^H)^T$.

Similarly, $t_1$ can be re-written as 

\vspace{-6mm}
\begin{eqnarray}
	t_1 = \boldsymbol{\theta}^H \mathbf v^{\ast} + \boldsymbol{\theta}^T \mathbf v,
\end{eqnarray}
\vspace{-5mm}

\noindent where $\mathbf v = \left[ \left(\mathbf D_2\right)_{1,1}, \cdots, \left(\mathbf D_2\right)_{N,N} \right]^T$, and \\
$\mathbf D_2 = (\alpha / \sigma_c^2) \mathbf G \mathbf W \mathbf W^H \mathbf F^H \mathbf H$. 
Thus, the sub-problem 2 can be concisely re-written as

\vspace{-6mm}
\begin{subequations} \label{eqn:sub_prob_2_new}
	\begin{eqnarray} 
		\!\!\!\!\!\!\!\!\!\max_{\boldsymbol{\theta}}&&	\!\!\!\!\!	f_1(\boldsymbol{\theta})=t_4(\boldsymbol{\theta})+t_2(\boldsymbol{\theta})+t_1(\boldsymbol{\theta})\label{eqn:obj_func_new} \\ \!\!\!\!\!\!\!\!\! \mathrm{s.t.}&& \!\!\!\!\! |{\boldsymbol\theta}_{n,1}| = 1,\;\; \forall n \in \{1,\cdots,N\} \label{eqn:unit_modu_new}
	\end{eqnarray}
\end{subequations}
\vspace{-6mm}

\noindent where ${\boldsymbol \theta}_{n,1}$ is the 
$n$-th element of the column vector $\boldsymbol{\theta}$.
The constraint in (\ref{eqn:unit_modu_new}) defines a complex circle manifold.  So, the second sub-problem has been re-formulated as an optimization problem with a quartic polynomial objective function constrained by an oblique manifold, which can be efficiently solved by a standard oblique manifold optimization method as follows.

\noindent {Let us define the oblique manifold based on the unit modulus constraint of (\ref{eqn:unit_modu_new})}

\vspace{-6mm}
\begin{eqnarray}
	\mathcal O = \{\boldsymbol{\theta} \in 
	\mathbb C^N | [\boldsymbol\theta \boldsymbol\theta^H]_{n,n}=1, \forall n=1, \cdots, N \}.
\end{eqnarray}  
\vspace{-6mm}

The tangent space of $\mathcal O$ at a certain point $\boldsymbol\theta_j \in 
\mathbb C^N$ can be written in a set form as

\vspace{-6mm}
\begin{eqnarray}
	T_{\boldsymbol\theta_j} \mathcal O = \{\mathbf x \in 
	\mathbb C^N | \textrm{Re}\{\mathbf x \circ \boldsymbol\theta_j^{\ast}\}=\mathbf 0_{N \times 1}\}.
\end{eqnarray}
\vspace{-6mm}

The Euclidean gradient of the re-arranged objective function $f_1(\boldsymbol{\theta})$ in (\ref{eqn:obj_func_new}) can be derived as

\vspace{-5mm}
\begin{small}
\begin{eqnarray} \label{eqn:euc_grad}
	\!\!\!\!\!\!\!\!\!\!\!\!\!\!\!\!\!\!&& \nabla f_1(\boldsymbol{\theta}) = (1-\alpha) \eta^2 / \sigma_r^2 \sum_{i=1}^{M} \sum_{j=1}^{M} [(\mathbf Z_{ij}+\mathbf Z_{ij}^T) \boldsymbol{\theta} \boldsymbol{\theta}^H \mathbf Z_{ij}^{\ast}\boldsymbol{\theta}^{\ast} \nonumber \\ \!\!\!\!\!\!\!\!\!\!\!\!\!\!\!\!\!\!&&+ (\mathbf Z_{ij}^{\ast}+\mathbf Z_{ij}^H) \boldsymbol{\theta}^{\ast} \boldsymbol{\theta}^T \mathbf Z_{ij}\boldsymbol{\theta}]   + 2\mathbf D_1 \boldsymbol{\theta} + 2 \mathbf v,
\end{eqnarray}
\end{small}
\vspace{-6mm}

\noindent where $\mathbf Z_{ij} = \mathbf R \circ (\mathbf G \mathbf w_j \mathbf g_i^T)^T$. Readers can refer to \cite{Alhujaili2020} for the derivation of gradient of $t_4(\boldsymbol{\theta})$. As we all know, the Euclidean gradient is the direction in which the objective function increases fastest. However, with the complex circle manifold constraint, we are not able to directly use the Euclidean gradient as the search direction, which should be in the tangent space of the manifold $\mathcal O$ at a certain point $\boldsymbol\theta_j$, i.e. $T_{\boldsymbol\theta_j} \mathcal O$. Thus we calculate the projection of the Euclidean gradient onto $T_{\boldsymbol\theta_j} \mathcal O$ instead as  search direction as follows \cite{Khaled2019}

\vspace{-6mm}
\begin{eqnarray} \label{eqn:rieman_grad}
	\mathrm{grad}_{\boldsymbol\theta_j}f_1 = \nabla f_1(\boldsymbol{\theta}_j) - \mathrm{Re}\{\nabla f_1(\boldsymbol{\theta}_j) \circ \boldsymbol{\theta}_j^{\ast}\}\circ \boldsymbol{\theta}_j.
\end{eqnarray}
\vspace{-6mm}

\noindent This is also known as the Riemannian gradient, which is a tangent vector of $\mathcal O$ at  $\boldsymbol{\theta}_j$, the direction in which the objective function increases fastest.

We denote the variable to be optimized $\boldsymbol{\theta}$ in the $j$-th iteration as $\boldsymbol{\theta}_j$. To find the value of $\boldsymbol{\theta}$ in the $(j+1)$-th iteration which is $\boldsymbol{\theta}_{j+1}$, we first let $\boldsymbol{\theta}_{j+1} = \boldsymbol{\theta}_{j} + \delta_j \mathrm{grad}_{\boldsymbol\theta_j}f_1$ where $\delta_j$ is the step size and $\mathrm{grad}_{\boldsymbol\theta_j}f_1$ is the search direction. In this case, $\boldsymbol{\theta}_{j+1}$ will be no longer on the complex circle manifold $\mathcal O$, so a retraction operation is needed which normalizes  $\boldsymbol{\theta}_{j+1}$ element-wisely to retract $\boldsymbol{\theta}_{j+1}$ back onto $\mathcal O$, i.e.

\vspace{-5mm}
\begin{eqnarray} \label{eqn:theta_update}
	\boldsymbol{\theta}_{j+1} = (\boldsymbol{\theta}_{j} + \delta_j \mathrm{grad}_{\boldsymbol\theta_j}f_1) \circ \frac{1}{|\boldsymbol{\theta}_{j} + \delta_j \mathrm{grad}_{\boldsymbol\theta_j}f_1|}.
\end{eqnarray}
\vspace{-4mm}

For the sake of exposition, the overall algorithm for solving the alternating optimization of radar precoder matrix $\mathbf W$ and IRS phase shift matrix $\mathbf \Theta$ is given in Algorithm \ref{alg:alt_opt}. In addition, $\varepsilon$ in Algorithm \ref{alg:alt_opt} is an error tolerance indicator. {Algorithm \ref{alg:alt_opt} is essentially the same with the methodology of \cite{Wang2021} which is manifold optimization based alternating optimization. However, our objective here is SNR maximization, which is a quartic function in $\mathbf \Theta$, while the objective of \cite{Wang2021} is beampattern deviation and multi-user interference minimization which is a quadratic function in $\mathbf \Theta$. } 

\vspace{-3mm}
\begin{algorithm}\label{alg:alt_opt}
	\SetAlgoLined
	\KwResult{Return $\mathbf W$ and  $\mathbf \Theta$.}
	\textbf{Initialization:} $\mathbf \Theta = \mathbf \Theta_0$, $\boldsymbol{\theta}_0=\mathbf\Theta_0 \mathbf 1_{N \times 1}$, $j=0$\;
	\While{ $\left(j<j_{\max}\;\&\&\;\big|\!\left[f_{1}^{(j)} \!-\! f_{1}^{(j-1)}\right]\!/\!f_{1}^{(j-1)}\!\big| > \varepsilon\right)$ }{
		Solve the problem of  (\ref{eqn:sub_prob_1}) for $\mathbf W$.
		
		Calculate  Euclidean gradient   $\nabla f_1(\boldsymbol{\theta}_j)$ based on (\ref{eqn:euc_grad});
		
		Compute Riemannian gradient $\mathrm{grad}_{\boldsymbol\theta_j}f_1$ as (\ref{eqn:rieman_grad});
		
		Update $\boldsymbol\theta_{j}$ to $\boldsymbol\theta_{j+1}$ as (\ref{eqn:theta_update});
		
		$j = j+1$;
		
		$\mathbf \Theta = \text{diag}(\boldsymbol\theta_{j+1})$;
		
	}
	\caption{Complex circle manifold constrained  iterative weighted received SNR maximization}
\end{algorithm}
\vspace{-4mm}

\vspace{-1mm}
\section{numerical results}
\vspace{-1mm}

\label{sec:typestyle}
We present  numerical results to demonstrate the convergence of the proposed  method, and quantify the advantages of the IRS-aided DFRC system. 
The channels $\mathbf F$ and $\mathbf H$ are simulated as flat Rayleigh fading  and $\mathbf G$ is Rician.

\begin{figure}[!t]\centering\vspace{-4mm}
	\includegraphics[width=0.35\textwidth]{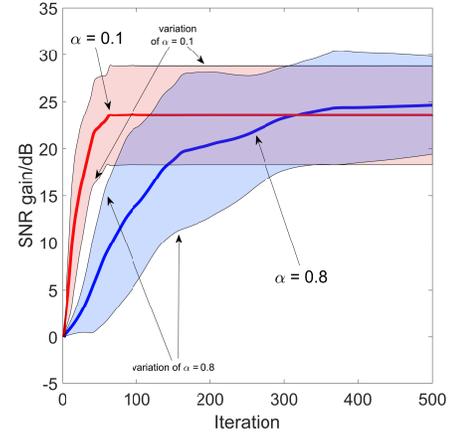}\vspace{-3 mm}
	\caption{Rate of convergence of the proposed alternating algorithm. System parameters: $P_0=30$ dBm, $M=8$, $N=64$.}
	\label{fig:Fig_convergence}\vspace{-5mm}
\end{figure}

{Fig. \ref{fig:Fig_convergence} shows the convergence rate of the proposed weighted sum received SNR maximization algorithm for different values of the weighting parameter ($\alpha$).   Solid lines indicate  the mean of SNR  gain  over 20 iterations, and the shaded area around the mean are bounded by the mean plus or minus standard deviation, indicating variance between different runs. It can be observed from Fig. \ref{fig:Fig_convergence} that a smaller $\alpha$ leads to faster convergence.  It should be noted that the radar SNR is a quartic function of the IRS phase shift matrix ($\mathbf \Theta$), while that for communication users is quadratic into $\mathbf \Theta$. Therefore the radar SNR is more sensitive to the change of $\mathbf \Theta$, and the radar SNR increases faster with regard to iteration number. Smaller $\alpha$ assigns larger weight for radar SNR, and the weighted SNR increases faster in this scenario.}

\begin{figure}[!t]\centering\vspace{-0mm}
	\includegraphics[width=0.35\textwidth]{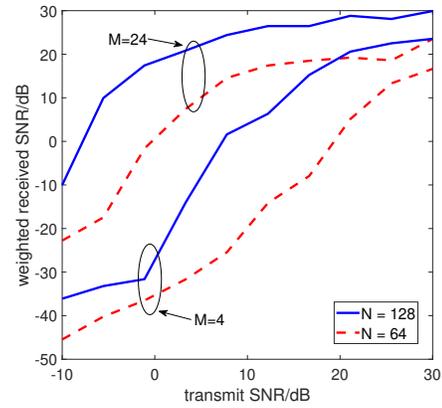}\vspace{-3 mm}
	\caption{Weighted received SNR versus transmit SNR at the radar. System parameters: $\alpha = 0.5$.}
	\label{fig:Fig_sum_rate}\vspace{-7mm}
\end{figure}

{In Fig. \ref{fig:Fig_sum_rate}, the relationship between the weighted received SNR and transmit SNR is displayed. The weighted received SNR can be boosted by increasing the number of radar antennas ($M$), the number of IRS elements ($N$), 
or the transmit SNR/power budget $P_0$.}
Therefore, the usefulness of IRS deployment for the DFRC system is quantified.

\begin{table}[]
	\begin{center}			
		\caption{Simulation Parameters}
		\vspace{-2mm}
		\label{tab:sys_para}
		\begin{tabular}{p{6.5cm}|p{1.0cm}} 
			\hline
			Parameter & Value
			\\\hline				
			Iterative algorithm error tolerance $\varepsilon$ (dB) & $- 30$
			\\\hline
			Maximum number of iterations $j_{\max}$ & 500
			\\\hline
			Iterative algorithm step size $\delta_j$ & $10^{-1}$
			\\\hline
			Number of communication receivers & 5
			\\\hline
			Rician factor of  $\mathbf G$ channel $K_G$ (dB) & 0
			\\\hline
			Radar receive array adjacent antenna distance & $0.5 \lambda$
			\\\hline
			Radar transmit array adjacent antenna distance&$0.5 \lambda$
			\\\hline
			Radar beampattern deviation threshold $\gamma_{bp}$ (dB)&$10$
			\\\hline
			Average noise power per radar receiver antenna   $\sigma_r^2$ (dB)&$0$
			\\\hline
			Average noise power per communication receiver antenna $\sigma_c^2$ (dB)&$0$
			\\\hline
		\end{tabular}
	\end{center}
	\vspace{-1mm}
\end{table}

\vspace{-2mm}
\section{conclusions}
\vspace{-1mm}
\label{sec:conclusions}
We have proposed a novel  IRS-aided DFRC system design and a manifold optimization based alternating optimization algorithm to maximize the weighted sum received SNR while considering the constant modulus constraint for IRS weights. The simulations have validated the convergence of the proposed optimization algorithm and  demonstrated the advantage of exploiting IRS in the DFRC system to improve the radar and communication performances. 

\clearpage


\balance


\vspace{12pt}

\bibliographystyle{IEEEtran}
\bibliography{IEEEabrv,References}

\begin{thebibliography}{10}
\providecommand{\url}[1]{#1}
\csname url@samestyle\endcsname
\providecommand{\newblock}{\relax}
\providecommand{\bibinfo}[2]{#2}
\providecommand{\BIBentrySTDinterwordspacing}{\spaceskip=0pt\relax}
\providecommand{\BIBentryALTinterwordstretchfactor}{4}
\providecommand{\BIBentryALTinterwordspacing}{\spaceskip=\fontdimen2\font plus
\BIBentryALTinterwordstretchfactor\fontdimen3\font minus
  \fontdimen4\font\relax}
\providecommand{\BIBforeignlanguage}[2]{{%
\expandafter\ifx\csname l@#1\endcsname\relax
\typeout{** WARNING: IEEEtran.bst: No hyphenation pattern has been}%
\typeout{** loaded for the language `#1'. Using the pattern for}%
\typeout{** the default language instead.}%
\else
\language=\csname l@#1\endcsname
\fi
#2}}
\providecommand{\BIBdecl}{\relax}
\BIBdecl

\bibitem{Liu2020}
F.~Liu, C.~Masouros, A.~P. Petropulu, H.~Griffiths, and L.~Hanzo, ``Joint radar
  and communication design: Applications, state-of-the-art, and the road
  ahead,'' \emph{{IEEE} Trans. Commun.}, vol.~68, no.~6, pp. 3834--3862, 2020.

\bibitem{Zhang2020d}
J.~A. Zhang, F.~Liu, C.~Masouros, R.~W. Heath~Jr, Z.~Feng, L.~Zheng, and
  A.~Petropulu, ``An overview of signal processing techniques for joint
  communication and radar sensing,'' \emph{arXiv preprint arXiv:2102.12780},
  2021.

\bibitem{Xu2021}
Z.~Xu and A.~Petropulu, ``A wideband dual function radar communication system
  with sparse array and {OFDM} waveforms,'' \emph{arXiv preprint
  arXiv:2106.05878}, 2021.

\bibitem{Liu2018}
F.~Liu, L.~Zhou, C.~Masouros, A.~Li, W.~Luo, and A.~Petropulu, ``Toward
  dual-functional radar-communication systems: Optimal waveform design,''
  \emph{{IEEE} Trans. Signal Process.}, vol.~66, no.~16, pp. 4264--4279, 2018.

\bibitem{Wu2020}
Q.~Wu and R.~Zhang, ``Towards smart and reconfigurable environment: Intelligent
  reflecting surface aided wireless network,'' \emph{{IEEE} Commun. Mag.},
  vol.~58, no.~1, pp. 106--112, 2020.

\bibitem{Wu2019}
------, ``Intelligent reflecting surface enhanced wireless network via joint
  active and passive beamforming,'' \emph{{IEEE} Trans. Wireless Commun.},
  vol.~18, no.~11, pp. 5394--5409, 2019.

\bibitem{Buzzi2021}
S.~Buzzi, E.~Grossi, M.~Lops, and L.~Venturino, ``Radar target detection aided
  by reconfigurable intelligent surfaces,'' \emph{{IEEE} Signal Process.
  Lett.}, vol.~28, pp. 1315--1319, 2021.

\bibitem{Lu2021}
W.~Lu, B.~Deng, Q.~Fang, X.~Wen, and S.~Peng, ``Intelligent reflecting
  surface-enhanced target detection in {MIMO} radar,'' \emph{IEEE Sensors
  Letters}, vol.~5, no.~2, pp. 1--4, 2021.

\bibitem{Lu2021a}
W.~Lu, Q.~Lin, N.~Song, Q.~Fang, X.~Hua, and B.~Deng, ``Target detection in
  intelligent reflecting surface aided distributed {MIMO} radar systems,''
  \emph{IEEE Sensors Letters}, vol.~5, no.~3, pp. 1--4, 2021.

\bibitem{Wang2021a}
X.~Wang, Z.~Fei, J.~Guo, Z.~Zheng, and B.~Li, ``{RIS}-assisted spectrum sharing
  between {MIMO} radar and {MU-MISO} communication systems,'' \emph{{IEEE}
  Wireless Commun. Lett.}, vol.~10, no.~3, pp. 594--598, 2021.

\bibitem{Aubry2021}
A.~Aubry, A.~De~Maio, and M.~Rosamilia, ``Reconfigurable intelligent surfaces
  for {N-LOS} radar surveillance,'' \emph{{IEEE} Trans. Veh. Technol.}, pp.
  1--1, 2021.

\bibitem{Wang2021b}
F.~Wang, H.~Li, and J.~Fang, ``Joint active and passive beamforming for
  {IRS}-assisted radar,'' \emph{{IEEE} Signal Process. Lett.}, pp. 1--1, 2021.

\bibitem{Wang2021}
X.~Wang, Z.~Fei, Z.~Zheng, and J.~Guo, ``Joint waveform design and passive
  beamforming for {RIS}-assisted dual-functional radar-communication system,''
  \emph{{IEEE} Trans. Veh. Technol.}, vol.~70, no.~5, pp. 5131--5136, 2021.

\bibitem{Jiang2021}
Z.-M. Jiang, M.~Rihan, P.~Zhang, L.~Huang, Q.~Deng, J.~Zhang, and E.~M.
  Mohamed, ``Intelligent reflecting surface aided dual-function radar and
  communication system,'' \emph{{IEEE} Syst. J.}, pp. 1--12, 2021.

\bibitem{Alhujaili2020}
K.~Alhujaili, V.~Monga, and M.~Rangaswamy, ``Quartic gradient descent for
  tractable radar slow-time ambiguity function shaping,'' \emph{IEEE
  Transactions on Aerospace and Electronic Systems}, vol.~56, no.~2, pp.
  1474--1489, 2020.

\bibitem{Khaled2019}
------, ``Transmit {MIMO} radar beampattern design via optimization on the
  complex circle manifold,'' \emph{{IEEE} Trans. Signal Process.}, vol.~67,
  no.~13, pp. 3561--3575, 2019.

\end{thebibliography}

\end{document}